\documentclass[preprint,onecolumn,amsmath,amssymb,aps,superscriptaddress]{revtex4-1}
\usepackage{graphicx}% Include figure files
\usepackage{dcolumn}% Align table columns on decimal point
\usepackage{bm}% bold math
\usepackage{color}
\usepackage{epstopdf}
\usepackage{amsmath}
\usepackage{amssymb}
\usepackage{array}
\usepackage{float}
\usepackage[normalem]{ulem}
\usepackage[export]{adjustbox}

\usepackage[colorlinks=true,citecolor=blue, linkcolor=blue, hyperfootnotes=false]{hyperref}

\makeatletter
\def\blfootnote{\xdef\@thefnmark{}\@footnotetext}
\makeatother

\begin{document}

\preprint{APS/123-QED}

\title{Minimum and maximum conductance of a thin film layer bridged interface: the role of anharmonicity and layer thickness}

\author{Jingjie Zhang}
\email{jz9wp@virginia.edu}
\affiliation{Electrical and Computer Engineering, University of Virginia, Charlottesville, Virginia 22904, USA}%

\author{Rouzbeh Rastgarkafshgarkolaei}
\email{In memory of Rouzbeh Rastgar.}
%\email{rr3ay@virginia.edu}
\affiliation{Mechanical and Aerospace Engineering, University of Virginia, Charlottesville, Virginia 22904, USA}%

\author{Carlos A. Polanco}
%\email{cap3fe@virginia.edu}
\affiliation{Materials Science and Technology Division, Oak Ridge National Laboratory, Oak Ridge, Tennessee 37831, USA}%

\author{Nam Q. Le}
%\email{nql6u@virginia.edu}
\affiliation{Research and Exploratory Development Department, The Johns Hopkins University Applied Physics Laboratory, Laurel, Maryland 20723, USA}%

\author{Keivan Esfarjani}
%\email{pamela@virginia.edu}
\affiliation{Mechanical and Aerospace Engineering, University of Virginia, Charlottesville, Virginia 22904, USA}%
\affiliation{Department of Physics, University of Virginia, Charlottesville, Virginia 22904, USA}%
\affiliation{Materials Science and Engineering, University of Virginia, Charlottesville, Virginia 22904, USA}%

\author{Pamela M. Norris}
%\email{pamela@virginia.edu}
\affiliation{Mechanical and Aerospace Engineering, University of Virginia, Charlottesville, Virginia 22904, USA}%

\author{Avik W. Ghosh}
\email{ag7rq@virginia.edu}
\affiliation{Electrical and Computer Engineering, University of Virginia, Charlottesville, Virginia 22904, USA}%
\affiliation{Department of Physics, University of Virginia, Charlottesville, Virginia 22904, USA}%

\date{\today}% It is always \today, today,
             %  but any date may be explicitly specified
\begin{abstract}
We study the role of anharmonicity at interfaces with an added intermediate layer designed to facilitate interfacial phonon transport. Our results demonstrate that while in the harmonic limit the bridge may lower the conductance due to fewer available channels, anharmonicity can strongly enhance the thermal conductance of the bridged structure due to added inelastic channels. Moreover, we show that the effect of anharmonicity on the conductance can be tuned by varying temperature or the bridge layer thickness, as both parameters change the total rate of occurrence of phonon-phonon scattering processes. Additionally, we show that the additive rule of thermal resistances (Ohm’s law) is valid for bridge layer thickness quite shorter than the average bulk MFP, beyond the regime it would be expected to fail.
\end{abstract}

\pacs{Valid PACS appear here}% PACS, the Physics and Astronomy
                             % Classification Scheme.
%\keywords{Suggested keywords}%Use showkeys class option if keyword
                              %display desired
\maketitle

%\tableofcontents
\section{Introduction}

The miniaturization of modern semiconductor devices to the nanoscale has led to a significant increase in heat density in integrated circuits \cite{Pop2010,Riedel2009,Park2016}. The accumulated heat is challenging to dissipate due to the elevated thermal resistance resulting from a large number of material interfaces. Currently, resistance at a single interface can account for up to 30-40\% of the total device thermal resistance, as in the case of the GaN/SiC interface in GaN high electron mobility transistors \cite{Sarua2007}. The resistance to heat dissipation caused by interfaces is an important bottleneck for further scaling of semiconductor devices. However, the existing gap in our fundamental understanding of heat transfer across single and multiple interfaces in nanostructures hinders the development of effective thermal management methodologies.~\cite{Cahill2003,Hopkins2013,Cahill2014}

\begin{figure}[hp]
	\centering
	\includegraphics[width=82mm]{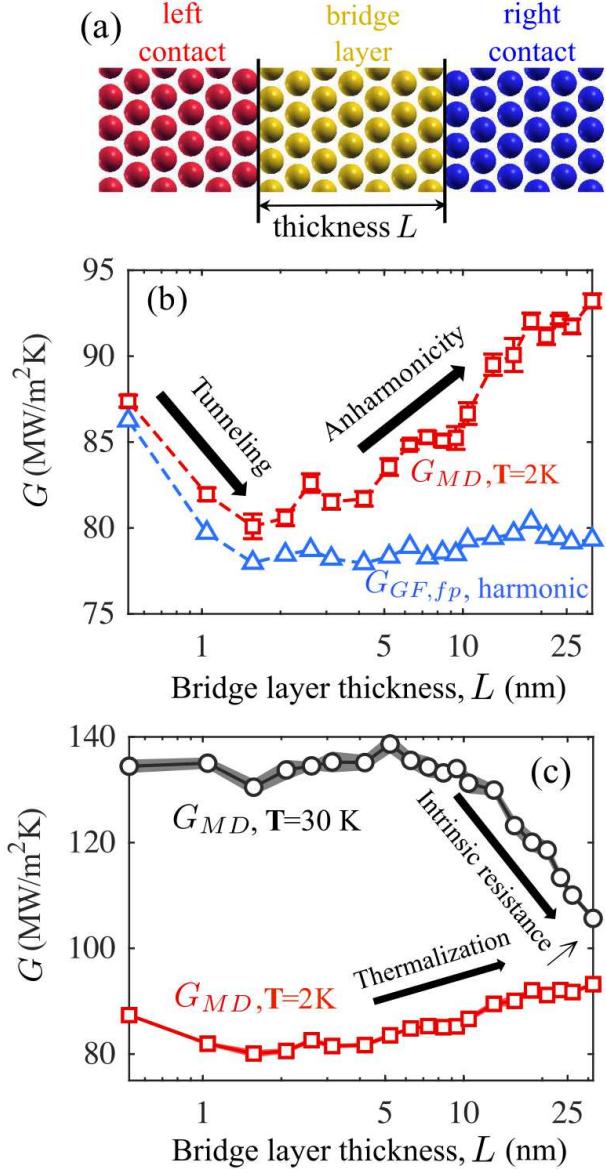}
	\caption{(a) Schematic of a bridged interface. The bridge layer mass is the geometric mean of masses of the materials at the interface $m_b=\sqrt{m_lm_r}$. This choice of the bridge layer mass can maximize the enhancement of the interfacial thermal conductance $G$ in the diffusive limit.\cite{Polanco2017,Rouzbeh2019} (b) The calculated interfacial thermal conductance in the harmonic limit,  $G_{GF,fp}$,(blue triangles) decreases to saturation with L, while conductance with weak anharmonicity $G_{MD}$, \textbf{T}=2 K shows a minimum. (c) The conductance with strong anharmonicity $G_{MD}$, \textbf{T}=30 K shows a maximum. The shaded areas in (c) denote the uncertainty based on 5 sets of NEMD simulations. We attribute the rise in the former (red squares) due to thermalization, and the drop in the latter (black circles) due to Umklapp scattering.}
	\label{fig_G}
\end{figure}

One promising approach to decreasing the thermal resistance at an interface is adding a bridging layer or intermediate thin film in between the interface (Fig.~\ref{fig_G}(a))~\cite{Gorham2014, Liang2011, Liang2012, English2012, Tian2012, Polanco2015}. This can effectively enhance interfacial thermal conductance by bridging either acoustic impedances (enhancing phonon transmission) or phonon frequency spectra (increasing the frequency range for phonon conduction by inelastic transport)\cite{English2012, Polanco2017, Liang2011, Lee2017, Zhang2018}. Nevertheless, the advantage of a bridging layer only exists in the additive regime, that is, when the total resistance can be treated as the sum of the resistances at the boundaries and the intrinsic resistance of the added layer. For instance, in the harmonic limit, where this assumption cannot be made, conductance is limited by the available modes that can conserve energy and transverse (parallel to the interface) momentum across the materials composing the bridged interface. As a result, the bridging layer, in the non-additive harmonic regime, decreases available modes and limits the possibility of enhancing conductance, $G$, for many combinations of materials.\cite{Polanco2015, Polanco2017, Rouzbeh2019} 

The transition from the non-additive to the additive regimes depends on anharmonic phonon-phonon scattering processes, and thus on the length of the intermediate layer, $L$, on the strength of anharmonicity $V_0$ (the third order of the interatomic force constants) , and on the temperature, \textbf{T}. A comprehensive study on how phonons flow across bridged interfaces in different transport regimes, accessible by varying these parameters, is still missing. Such a study could clarify how different scattering processes determine the transition between additive and non-additive regimes and thus enable better thermal engineering of devices.

The critical length scale at which the additive regime is valid is of significant importance for measurements of thermal conductivity of thin films, typically sandwiched between two bulk materials. Extraction of the thermal conductivities of these thin films, e.g. by time-domain thermoreflectance (TDTR) or Raman spectroscopy\cite{Jiang2016, Cahill2004,Calizo2007,Cai2010}, normally relies on the assumption that the total resistance can be treated as a summation of resistances in series. The validity of summing resistances is usually determined by comparing the bulk mean free paths (MFPs) with the thin film thicknesses. In many cases, however, the comparison is difficult, as thicknesses of the thin films are often on the order of nanometers while the bulk phonon MFPs span a wide range of scales from nanometers to micrometers. It is debatable whether the comparison between the bulk MFPs and the thin film thickness is an accurate measure to determine the transport regimes. In this paper, we vary the thin film thickness under different temperatures to determine the length scales at which the additive regime is valid. These length scales are then compared with the bulk MFPs to validate the role of bulk MFPs in determining the various transport regimes.

In this work, the thermal transport across a prototype model argon--bridge--``heavy argon'' interface (Fig.~\ref{fig_G}(a)) is investigated, with varying the layer thickness, $L$, and temperature, \textbf{T}. Our results demonstrate the existence of a minimum interfacial thermal conductance, $G$, with $L$ when anharmonicity is weak and the existence of a maximum $G$ with $L$ when anharmonicity is strong (Sec. ~\ref{sec_G}). The minimum thermal conductance appears at a short $L$ and is due to the competing roles of phonon tunneling and thermalization. The maximum thermal conductance is a result of two competing effects of anharmonicity: thermalization and Umklapp scattering.  Moreover, we show that the effect of anharmonicity on the conductance can be tuned by varying temperature or intermediate layer thickness, as both parameters can change the number of phonon-phonon scattering processes. In Section ~\ref{sec_lambda}, we study the critical length, $L_s$, at which the total resistance can be separated into components as resistances in series. Additionally, we compare the critial length to the bulk MFPs $\lambda_b$, and demonstrate that $L_s$ is much smaller than $\lambda_b$, suggesting that the additive regime can be extended to much smaller length scales than the bulk mean free path. 

\section{Methodology}
\label{sec_method}

Figure 1(a) depicts the system studied in this paper, a ``bridged interface.'' The left and right contacts as well as the bridge layer share the same face-centered cubic (fcc) crystal structure with interatomic interactions given by the same Lennard-Jones potential. The boundaries between adjacent materials are abrupt and clean without any lattice mismatch or defects. The atomic masses of the left and right materials are  $m_l = 40$ amu and  $m_r = 120$ amu respectively, and the atomic mass of the bridge layer is the geometric mean $m_b=\sqrt{m_rm_l}$ of those two. This choice of $m_b$ maximizes the enhancement of the conductance by a bridge layer in the diffusive regime.\cite{Polanco2017,Rouzbeh2019} The dissimilar atomic masses cause different vibrational properties in those materials. To study the dependence of conductance on the intermediate layer thickness and to determine the critical length $L_s$ to separate the resistances, the bridge layer thickness is varied from 1 to 60 conventional unit cells (lattice constant $a$ is 0.522 nm), and the temperature \textbf{T} is set to 0 K, 2 K or 30 K. Further details of the system are provided in Appendix~\ref{app:details}).

By changing temperature, \textbf{T}, we calculate thermal conductance across the system in three regimes: without anharmonicity (\textbf{T}=0 K), with weak anharmonicity (\textbf{T}=2 K), and with strong anharmonicity (\textbf{T}=30 K). Conductance at \textbf{T}=0 K, without any anharmonic interactions, is computed using the Landauer formalism 
\begin{equation}
G_{GF}=\frac{1}{A}\int\limits_0^\infty{\frac{\hbar\omega}{2\pi}\frac{\partial N}{\partial \textbf{T}}M\bar{T}d\omega}\xrightarrow[\text{limit}]{\text{classical}}\frac{k_B}{2\pi A}\int\limits_0^\infty{M\bar{T}d\omega},
\label{equG}
\end{equation}
where $A$ is the cross-sectional area, $\hbar$ is the reduced Planck constant, $N$ is the Bose--Einstein distribution, $k_B$ is the Boltzmann constant, $M$ is the number of available propagating modes and $\bar{T}$ is the average transmission per mode. $M\bar{T}$ is determined using Non-Equilibrium Green's Function (NEGF) by $M\bar{T}=\text{Trace}[\Gamma_l{\bf{G}}\Gamma_r{\bf{G}}^{\dagger}]$, with ${\bf{G}}$ the retarded Green's function and $\Gamma_{l}$ ($\Gamma_{r}$) the broadening matrix describing the interactions between the device---in this case, the bridge layer---and the left (right) contact material. In a system preserving symmetry across interfaces, the number of propogating modes equals to minimum modes conserving transverse momentum in all composing materials: $M(\omega)=\underset{k_\perp}{\sum}\underset{\alpha}{\text{min}}[M_\alpha(\omega,k_\perp)]$, where $\alpha$ is left/bridge/right material in this study. The number of modes in each individual bulk material ($M_{l/b/r}$ left, bridge and right materials) can be determined using the same method for calculating $M\bar{T}$ but with the contacts and device chosen as the same material; in this case, the transmission $\bar{T}$ is unity for each mode. Conductance including weak and strong anharmonicity is computed using Non-Equilibrium Molecular Dynamics (NEMD) at \textbf{T}=2 K and \textbf{T}=30 K respectively. Details of our NEGF and NEMD simulations are presented in Appendix~\ref{app:details}, including tests checking for domain size effects. ($G_{GF}$ in this paper denotes the harmonic conductance without any anharmonic interactions, but in principle NEGF can include anharmonic interactions despite the simulation is computationally expensive.\cite{Luisier2012, Mingo2006})
To calibrate the conductances from NEGF ($G_{GF}$) and NEMD ($G_{MD}$), the high temperature limit of Eq.~\ref{equG} is used in NEGF calculations ($\hbar\omega_{max}\ll k_B\textbf{T}$, with $\omega_{max}$ the maximum vibrational frequency of the system), so that phonons across the whole spectrum contribute equally to transport as in the classical limit. Furthermore, the contact resistance is excluded from $G_{GF}$ using
\begin{equation}
G_{GF,fp}=G_{GF}\frac{\Delta \textbf{T}_c}{\Delta \textbf{T}_i},
\label{equG_T}
\end{equation} 
where $\Delta \textbf{T}_i$ and $\Delta \textbf{T}_c$ are temperature differences at the interface and between the contact baths in NEMD simulations at \textbf{T}=2 K, as illustrated in Fig.~\ref{fig_appB_Tfile}(a) of the Appendix.~\ref{app:Rcontact}. In this way, the two-probe conductance measurement from NEGF is converted to a four-probe measurement that captures only temperature drops at the interface. The four probe conductance in NEGF without using the temperature differences from NEMD is provided in Appendix~\ref{app:Rcontact}.

\section{Minimum and maximum conductance versus bridge layer thickness}
\label{sec_G}

Depending on the anharmonic scattering rates (as controlled by temperature), conductance across the bridged interface exhibits different trends as the thickness of the intermediate layer increases (Figs.~\ref{fig_G}(b) and \ref{fig_G}(c)). With zero anharmonicity, $G_{GF,fp}$ initially decreases and quickly saturates at $\sim2$ nm (Fig.~\ref{fig_G}(b)). With weak anharmonicity, an initial decrease of $G_{MD}$ at \textbf{T}=2 K is followed by an upward trend (Fig.~\ref{fig_G}(b)), resulting in a local minimum in conductance with respect to bridge layer thickness. Finally with strong anharmonicity, $G_{MD}$ at \textbf{T}=30 K decreases after $\sim5$ nm following Fourier's law (Fig.~\ref{fig_G}(c)). {\iffalse{Although before $5$ nm the uncertainty of our simulations disallow strong claims, it seems that $G_{MD}$ at \textbf{T}=30 K presents an initial decrease followed by and increase, similar to those of $G_{MD}$ at \textbf{T}=2 K.}\fi} In this section, each trend is explained in terms of three different transport mechanisms: phonon tunneling, thermalization processes, and intrinsic resistance by Umklapp scattering.

Phonon tunneling explains the decreasing trend of $G$ versus $L$ in the harmonic limit or at low temperature (\textbf{T}=2 K) in  Fig.~\ref{fig_G}(b). By ``phonon tunneling,'' we refer to a non-zero, elastic phonon transmission across a bridged interface via an evanescent vibrational wave in the intermediate (bridge) layer. Contrary to propagating waves, i.e. eigenvectors of the harmonic equation of motion whose amplitude are constant along the crystal (normal phonons), evanescent waves decay exponentially in the crystal and thus cannot carry heat over long distances. Nevertheless, for layer thicknesses shorter than the decay length, evanescent waves can bridge propagating waves or phonons across two materials.\cite{Altfeder2010,Tian2010} 

\begin{center}
\begin{figure}[htp]
\includegraphics[width=87mm]{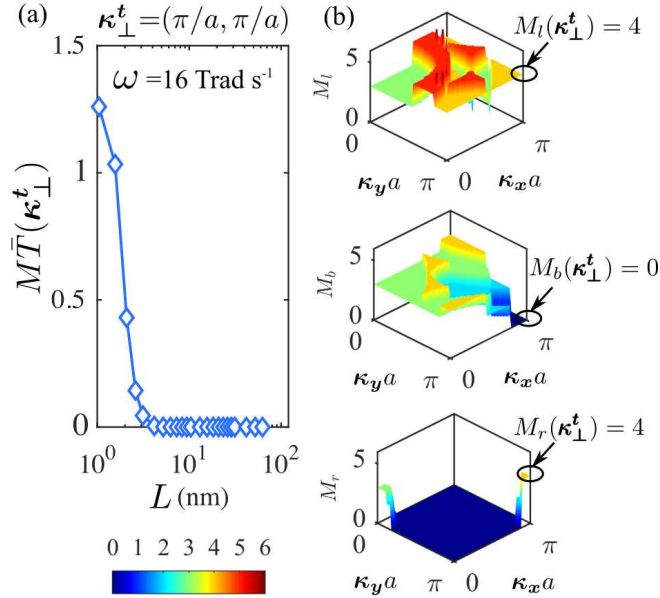}
\caption{At frequency $\omega$=1.6 Trad s$^{-1}$, (a) transmission for modes at {\boldmath$\kappa_{\perp}^t$}=($\pi/a$, $\pi/a$) is a monotonically decaying function of $L$, where $a=0.522$ nm is the conventional unit cell lattice constant; (b) number of modes in the left ($M_l$), bridge layer ($M_b$) and right ($M_r$) material respectively, showing an absence of modes in the bridging layer (black circles) leading to tunneling.}
\label{fig_tunneling}
\end{figure}
\end{center}

Phonon tunneling can be unambiguously identified in the harmonic limit, where nonzero transmission across the bridged interface {\it is only possible if} phonons conserve energy and transverse wavevector {\boldmath$\kappa_\perp$} in the plane of boundaries. Conservation of {\boldmath$\kappa_\perp$} results from the transverse symmetry of the abrupt material boundaries of our system being free from impurities, defects, lattice mismatch or interatomic mixing. Phonon transport across such boundaries is not acted upon by any forces in the transverse directions, and hence does not have any momentum (velocity) scattering in that direction.

Evidence of phonon tunneling in our system is given in Figure~\ref{fig_tunneling}. Figure~\ref{fig_tunneling}(a) shows $M\bar{T}$ versus $L$ for phonons at frequency $\omega$=16 Trad s$^{-1}$ and transverse wavevector {\boldmath$\kappa^t_\perp$}=$(\pi/a,\pi/a)$. $M\bar{T}$ arises from phonon tuneling because there are available propagating modes or phonons at {\boldmath$\kappa^t_\perp$}=$(\pi/a,\pi/a)$ only in the left and right contact materials but not in the bridge layer (see circled regions in Fig.~\ref{fig_tunneling}(b)). Thus vibrational energy transport across the bridge is only possible via evanescent modes. Also, $M\bar{T}$ decreases monotonically with length, as expected for heat-carrying evanescent waves. At {\boldmath$\kappa^t_\perp$}=$(\pi/a,\pi/a)$ there are similar phonon tunnelling contributions to conductance for frequencies between 15 and 16 Trad s$^{-1}$, where propagating modes are available only in the contacts but not in the bridge (see Fig.~\ref{projected_dispersion} in Appendix~\ref{app:figures}). 

Phonon tunneling sets in when the bridge modes at a given ($\omega$,$\kappa_\perp$) fall significantly below the mode counts in the contacts, not just when the former is zero. The sum of the decaying transmission of all evanescent vibrations in the bridge material results in a decreasing trend of $M\bar{T}$ and thus of $G_{GF}$ at short $L$ (Fig.~\ref{fig_G}(b)). When the contribution to $M\bar{T}$ from phonon tunneling becomes negligible, $M\bar{T}$ and $G_{GF}$ saturate because the Fabry-Perot oscillations in the transmission due to wave interference are partially destroyed by the summation over phonons with different wavelengths and then further averaged out by the integral over frequency.

Phonon tunneling and the initial decrease in $G_{GF}$ vs. $L$ can also be explained from another equivalent point of view similar to metal-induced gap states (MIGS) in electron transport, using local density of states (LDOS). When $L$ is very short, the LDOS in the bridge layer is permeated with levels from the contact materials that are not present in the DOS of the bulk bridge material, which may allow phonon transport across the interface. However, as $L$ increases, the LDOS of atoms in the bridge layer away from the boundaries recovers the DOS of the bulk bridge material, and the extra transport levels assisting transport disappear. This example emphasizes the importance of interfacial eigenmodes when considering transport.\cite{Gordiz2016sr,Gordiz2016JAPa,Gordiz2016JAP} 

\begin{figure*}[!ht]
	\centering
    \includegraphics[width=\textwidth]{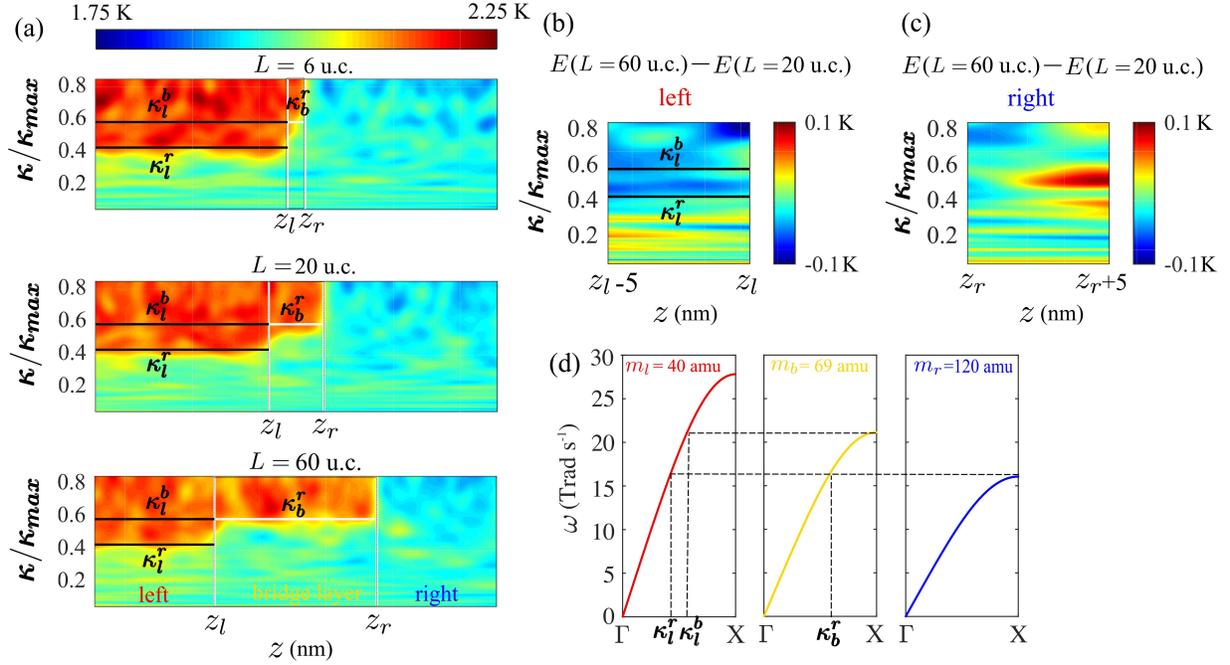}
	\caption{(a) The kinetic energy density distribution of longitudinal modes along $\langle$001$\rangle$ directions when $L$ is 6 u.c., 20 u.c. and 60 u.c. respectively in the ($z$, wavevectors) plane.} Modes with wavevector $\pmb{\kappa_l^r}$ in left and $\pmb{\kappa_b^r}$ in bridge layer material have the same frequency as the cut-off frequency in the heavy material (right contact material). $z_l$ and $z_r$ denote the location of the left and right boundaries. Phonons above the cut-off frequency of the heavy material accumulate in the left and in the bridge materials at 2 K. (b, c) The energy density difference between 60 u.c. and 20 u.c. systems at the 5 nm regions in the left and right materials close to the left boundary ($z_l$-5 nm to $z_l$) and right boundary ($z_r$ to $z_r$+5 nm), showing signature of thermalizaiton where high frequency phonons on the left scatter to low frequency regimes and transport across the interface. (d) Dispersion curves of the longitudinal phonon branches in the three different materials and relationship with the critical wavevectors, $\pmb{\kappa_l^r}$, $\pmb{\kappa_l^b}$, and $\pmb{\kappa_b^r}$.
\label{fig_wavelet}
\end{figure*}

In the case of weak anharmonicity (\textbf{T} = 2 K), the conductance initially decreases with the bridge layer thickness just as in the harmonic limit, but then deviates and increases with $L$ (Fig.~\ref{fig_G}(b)). This increase results from thermalization processes enabled by the weak anharmonicity in the system, as supported by analyzing the variations in energy density with respect to both space and wavevector. Distributions of the kinetic energy density in the NEMD simulations at steady state were calculated as functions of $z$ and $\pmb{\kappa}$ using the wavelet transform as in previous work~\cite{Nam2017} and plotted in Fig.~\ref{fig_wavelet}. For ease of interpretation, the energy density is reported in terms of the equivalent temperature according to the equipartition principle, given that the MD simulations are classical. The energy densities in Fig.~\ref{fig_wavelet} are therefore equivalent to temperatures near the average system temperature \textbf{T} = 2 K. For brevity, only the density distributions of longitudinal $\langle 001 \rangle$ modes are shown. The densities can also be plotted for the transverse modes and show similar behaviors, albeit at correspondingly lower cut-off frequencies than the longitudinal modes.

Across a range of bridge thicknesses from $L$ = 6 u.c. to 60 u.c., there is a significant excess of kinetic energy density in high-wavenumber modes in the light material on the left side (Fig.~\ref{fig_wavelet}(a)). The modes with excess energy are those with frequencies higher than the cut-off frequency of the heavy material on the right side; this corresponds to a sharp transition at the corresponding wavenumber $\pmb{\kappa_l^r}$. The relationships among these critical wavenumbers are shown in Fig.~\ref{fig_wavelet}(d). Likewise, in the middle bridging material with intermediate mass, there is excess energy density above the corresponding wavenumber $\pmb{\kappa_b^r}$. These distributions imply that phonon transmission across interfaces is predominantly elastic (frequency-preserving) at \textbf{T} = 2 K, and phonons with frequencies above the cut-off frequency of the heavy material are primarily reflected.

Closer inspection of the wavelet spectra also suggests a mechanism for the effect of increasing the bridging layer thickness. Specifically, we examined the energy distributions in the systems with the ``short'' $L = 20$ u.c.~bridging layer and the ``long'' $L = 60$ u.c.~bridging layer and compared the spectra in analogous spatial regions. In Fig.~\ref{fig_wavelet}(b), we show the difference in energy densities within the region 5 unit cells to the \emph{left} of the interface ($z_l$), $z \in [z_l - 5, z_l]$. While both systems exhibited excess energy density in the high-wavenumber modes $\pmb{\kappa} > \pmb{\kappa_l^r}$, as already seen in Fig.~\ref{fig_wavelet}(a), the magnitude of excess energy diminishes with increasing bridge layer thickness, resulting in $E(L = \mathrm{60~u.c.}) < E(L = \mathrm{20~u.c.})$ in the same high-wavenumber modes. Therefore, the increased bridge layer thickness correlates with more efficient thermalization of energy in the lighter material, while also correlating with increased total conductance. In Fig.~\ref{fig_wavelet}(c), the difference in energy densities is shown within the analogous region 5 unit cells to the \emph{right} of the interface $z_r$, $z \in [z_r , z_r+5]$. In this region, the energy density is greater on average in the system with the long $L = 60$ u.c.~bridging layer across all frequencies. Taken together, Figs.~\ref{fig_wavelet}(b,c) suggest a mechanism for increasing conductance with layer thickness $L$ due to weak anharmonic interactions scattering high frequency phonons with low or no transmission to frequencies below $\omega_r^{cut}$ where transmission is higher.

\begin{figure}[htp]
	\centering
	\includegraphics[width=86mm]{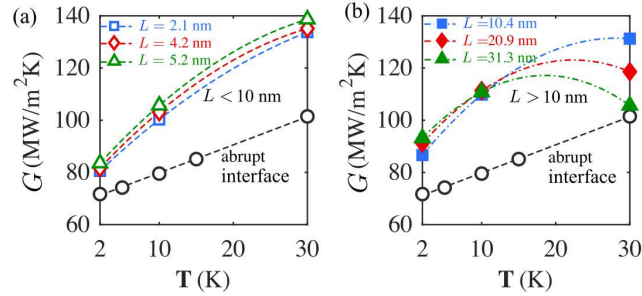}
	\caption{(a) At short length, conductance $G$ increases with \textbf{T} with a larger slope than that of an abrupt interface. Meanwhile, the longer the bridge layer thickness, the larger the conductance and the slope. These increases are due to thermalization. (b) When the bridge layer thickness is large enough, increasing \textbf{T} tends to decrease the conductance due to Umklapp scattering.}
	\label{fig_G_vs_T}
\end{figure}

That thermalization processes initially drive the increase in conductance with bridge layer thickness $L$ can also be verified by comparing the slopes of $G$ versus $\textbf{T}$ for systems with different layer thicknesses (Fig.~\ref{fig_G_vs_T}(a)). Increasing $\textbf{T}$ increases the rates of anharmonic phonon scattering and also tends to create a linear increase in conductance.\cite{Nam2017, Wu2014} Larger slopes indicate this kind of scattering-assisted transport enhancement is larger. Compared to the abrupt interface, the bridged interface conductance $G$ increases with $\textbf{T}$ with a larger slope, and this slope increases with $L$. Accordingly, we conclude that increasing the bridging layer thickness introduces more thermalization processes and thus increases the conductance.

As temperature $\textbf{T}$ rises further and the bridge layer thickness $L$ increases, the conductance of the bridged interfaces ultimately decreases (Fig.~\ref{fig_G_vs_T}(b) and Fig.~\ref{fig_G}(c)) at 30 K.  This happens when strong anharmonicity is present, and arises from Umklapp back scattering processes, where phonons moving in the transport direction are scattered to phonons with opposite velocity. We thus conclude that weak anharmonicity can effectively increase phonon transport across moderately thick bridging interfaces by increasing thermalization and inelastic transport modes, while strong anharmonicity can reduce phonon transport across thick interfaces by increasing resistive scattering. 

Our results show the bridge layer thickness $L$ can be used as another parameter to tune the strength of anharmonicity in addition to temperature \textbf{T}. The effects of these two parameters can be explained by the Fermi's Golden Rule. Take the three-phonon decay scattering rate derived from Fermi's golden rule as an example:
\begin{equation}
    \Gamma^{-}_{j}=\frac{1}{N}\sum_{{j^{\prime}}{j^{\prime\prime}}}\frac{\hbar\pi}{4}\frac{N^{\prime}_0+N^{\prime\prime}_0+1}{\omega_{j}\omega_{j^{\prime}}\omega_{j^{\prime\prime}}}|V^{-}_{jj^{\prime}j^{\prime\prime}}|^2\delta(\omega_{j}-\omega_{j^{\prime}}-\omega_{j^{\prime\prime}}),
\end{equation}
where $N^{\prime}_0$ (similar for $N^{\prime\prime}_0$) is the Bose-Einstein occupancy for mode $\omega_{j^{\prime}}$ (or mode $\omega_{j^{\prime\prime}}$) and $V^{-}_{jj^{\prime}j^{\prime\prime}}$ is the three-phonon anharmonic scattering matrix element relating modes $j,j^{\prime}$ and $j^{\prime\prime}$.\cite{Tian2012b,Li2014} Increasing the bridge layer thickness $L$ increases the space for phonons to interact with each other, i.e, increases the scattering phase space $\sum_{{j^{\prime}}{j^{\prime\prime}}}\delta(\omega_{j}-\omega_{j^{\prime}}-\omega_{j^{\prime\prime}})$. Meanwhile, increasing the temperature \textbf{T} increases the displacement of atoms, which increases the occupation of phonons  $N^{\prime}_0+N^{\prime\prime}_0+1$. 

\section{Length scale to separate resistance \texorpdfstring{$L_s$}[ versus mean free path \texorpdfstring{$\lambda_b$}[ }
\label{sec_lambda}

\begin{figure}[htp]
	\centering
	\includegraphics[width=86mm]{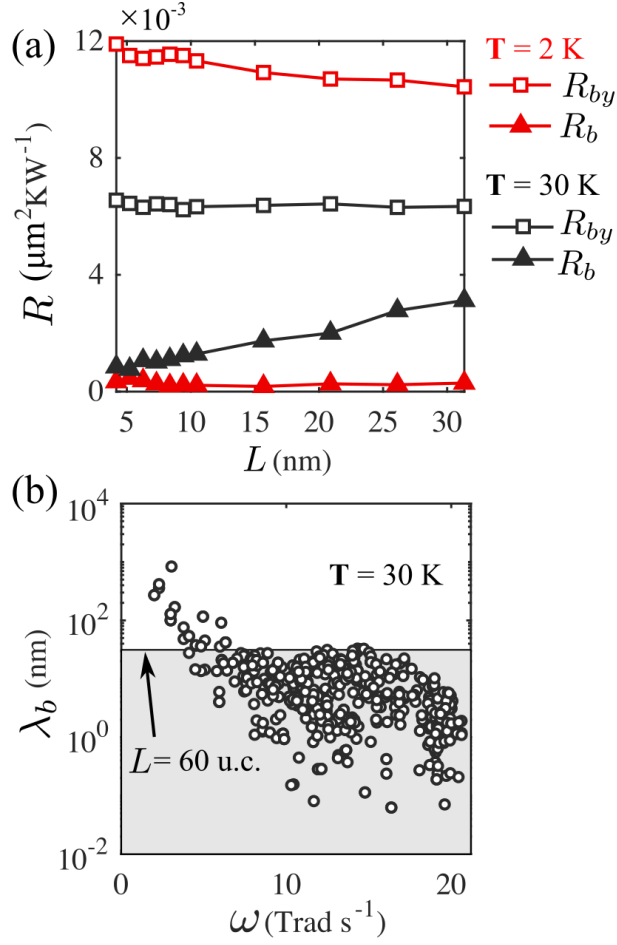}
	\caption{(a) The thermal boundary resistance $R_{by}$ (open square) and the thermal resistance in the bridge layer $R_b$ (solid up-triangle) at 2 K (red) and 30 K (black); Note the linear increase of $R_b$ versus $L$ and the constancy of boundary resistances indicating the Ohmic behavior at 30 K. (b)Bulk mean free path of the bridge layer $\lambda_b$ from normal mode decomposition. A quite small percentage of the modes, which have frequencies below 5 Trad s$^{-1}$ are expected to propagate ballistically while the rest follow a diffusive process.}
	\label{fig_R}
\end{figure} 

Boundaries and the bridge layer of a bridged interface system contribute to the total thermal resistance ($R_{tot}=1/G$) in two different ways. In the ballistic transport regime, phonons transport through the bridge layer without any backscatterings, thus the major contribution to the resistance stems from the boundaries. On the other hand, if the system is in the diffusive regime, both boundaries and the bridge layer contribute to the total resistance, and their resistances can be summed together. In addition, the bridge layer resistance behaves as an intrinsic resistor following Fourier's law, while the boundary resistances should be independent of each other and of the bridge layer thickness $L$.

We quantified the boundary resistances ($R_{l,by}$,$R_{r,by}$ are resistances for the left and right boundary respectively) and the intrinsic resistance of the bridge layer ($R_b$) as the ratio of temperature drop over heat flux, such that:
\begin{equation}\label{eq_R}
  \begin{array}{l}
    R_{l/r,by}=\Delta\textbf{T}_{l/r}/q=[(\Delta\textbf{T}_{l/r}/\Delta\textbf{T}_i]R_{tot}\\
    R_{b}=\Delta\textbf{T}_b/q=[\Delta\textbf{T}_b/\Delta\textbf{T}_i]R_{tot}
  \end{array}
\end{equation}
where $\Delta\textbf{T}_l$ and $\Delta\textbf{T}_r$ are the temperature drops at the left and right boundary respectively, $\Delta\textbf{T}_b$ is the temperature drop within the bridge layer and $\Delta\textbf{T}_i$ is the total temperature drop across the bridged interface (as illustrated in Fig.~\ref{fig_appB_Tfile}(b) in Appendix.~\ref{app:Rcontact}). 

The near zero value of $R_b$ at 2 K indicates the majority of phonons transport ballistically through the bridge layer, and all scattering events leading to resistance happen at the boundaries (Fig.~\ref{fig_R}(a)). $R_{by}$ ($R_{by}$=$R_{l,by}$+$R_{r,by}$) decreases with $L$, in agreement with the previous discussions on $G$ versus $L$, suggesting that all benefits on the conduction by thermalization processes are at the boundaries. The ballistic feature of $R_b\sim0$ $\mu$m$^2$KW$^{-1}$ also suggests that the resistance cannot simply be treated as resistances in series at 2 K.

At 30 K, the trends of resistances versus $L$ reverse, indicating that bridge interfaces with layer thickness larger than  $\sim$5 nm are in a diffusive transport regime, and the resistances can be treated as resistances in series. This is supported by the following phenomena. First, $R_{by}$ does not change with $L$, suggesting $R_{by}$ is not influenced by the bridge layer thickness. Second, comparing the bulk mean free path of the bridge layer material $\lambda_b$ at 30 K (Fig.~\ref{fig_R}(b)) with the bridge layer thickness $L$ ($\lambda_b$ is computed by the normal mode decomposition technique explained in Appendix.~\ref{app:NMD}), only a few phonons (with $\omega\le$ 5 Trad s$^{-1}$) have bulk mean free paths longer than 60 u.c. ($L$=31.3 nm). These phonons only contribute to 6.14\% of the total thermal conductance at 30 K, this indicates the $L$=60 u.c. system is almost in the diffusive limit, and $R_{by}$ from $\sim$5 nm is the same as $R_{by}$ in the diffusive limit ($R_{by}$ at 60 u.c.). Last but not the least, the intrinsic resistance of the bridge layer $R_b$ increases linearly with $L$, demonstrating a behavior following the Fourier's law. Note that $R_{by}$ at 5 nm is the same value as $R_{by}$ in the $L$=60 u.c. (31.3 nm) system. This indicates that the resistances can be treated as resistances in series from a very short length scale($\sim$5 nm). The majority of phonons have $\lambda_b$ longer than 5 nm at 30 K. Therefore, the common criterion for diffusive transport $\lambda_b<L$ appears to be stricter than necessary. The reason could be that the bulk junction material MFP does not take the interface scattering, which can be quite large, into consideration.

\section{Conclusion}
We studied the role of temperature and the bridging layer thickness on the thermal conductance of a bridged interface. Our results demonstrate the existence of minimum and maximum conductance by varying either temperature or layer thickness. These phenomena are due to dual roles of anharmoncity that it can either enhance or hinder phonon transport. The minimum thermal conductance is a result of ``phonon tunneling'' and thermalization effects with weak anharmonicity. The maximum thermal conductance is due to additional Umklapp scattering with strong anharmonicty. Furthermore, we demonstrated the summation over thermal resistance rule can be used at a much shorter layer thickness than the bulk mean free path of the intermediate layer, indicating the comparison between bulk mean free path and intermediate layer thickness is too strict a rule to determine transport regimes for thin films bridging two materials. 

\begin{acknowledgments}
J.Z. and A.W.G acknowledge the support from ``Graduate opportunity (GO!)" program associated with Center for Nanophase Materials Sciences (CNMS) at Oak Ridge National Laboratory (ORNL). R.R and P.M.N. acknowledge the financial support of the Air Force Office of Scientific Research (Grant No. FA9550-14-1-0395). N.Q.L. acknowledges support from the U.S. Naval Laboratory (NRL) through the National Research Council Research Associateship Programs. C.A.P. acknowledges support from the Laboratory Directed Research and Development Program of Oak Ridge National Laboratory, managed by UT-Battelle, LLC, for the U.S. Department of Energy. Computational work was performed using resources of the Advanced Research Computing Services at the University of Virginia and the ``Campus Compute Co-operative (CCC)"~\cite{Grimshaw2016}. The authors are grateful for useful discussions with LeighAnn Larkin.\\
J.Z. and R.R. contributed equally to this work.

\end{acknowledgments}

\appendix
\section{Simulation Details}
\label{app:details}
We study thermal transport across bridged interfaces shown in Fig. 1(a). In this study, all material properties (interatomic potentials, crystal structures and lattice constant) except the atomic masses stay invariant throughout the whole system. The crystal structure for the three components is face-centered cubic with one atom per primitive unit cell, and the lattice constant $a$ is 0.522 nm. Interfaces are abrupt, free of defects and without lattice mismatch. The atomic mass in the bridge layer is the geometric mean of the contact masses $m_b=\sqrt{m_lm_r}$ ($m_l$=40 amu, $m_r$=120 amu and $m_b$=69.28 amu), thus its impedance and vibrational spectrum bridge those of the contacts. The bridge layer thickness $L$ is varied from 1 u.c. to 60 u.c., and the ambient temperature is set to either 0 K, 2 K or 30 K.

The Lennard-Jones (LJ) potential is used to describe the interatomic interactions: $U_{LJ}(r_{ij}) = 4\epsilon[(\sigma/r_{ij} )^{12} - (\sigma/r_{ij} )^6]$, with parameters $\epsilon = 0.0503$ eV and $\sigma = 3.37$ \AA. These parameters are identical to those in our previous works.~\cite{Polanco2017, Rouzbeh2019} The cut-off distance for the potential is 2.5$\sigma$, which includes atomic interactions up to $5^{th}$ nearest atomic neighbors. In the harmonic Green's function calculations, the interatomic force constants come from the $2^{nd}$ order expansion of the LJ potential, and the interactions also include up to $5^{th}$ nearest atomic neighbors. To benchmark the two methodologies used to compute conductance, NEMD and NEGF simulations, we calculate the conductance of an abrupt interface, without the bridge layer. The conductance of such a system given by NEMD at 2 K, $G_{MD}$ is 71.71$\pm$0.36 MWm$^{-2}$K$^{-1}$, and given by NEGF without contact resistances (by approach (b) in Appendix~\ref{app:Rcontact}) in the classic limit is 70.14 MWm$^{-2}$K$^{-1}$. These values show excellent agreement between each other, allowing us to compare results from these two methods. 

NEGF simulations are performed with 200 grid point sampling the frequency interval from 0--40 Trad s$^{-1}$ and a 100$\times$100 wavevector mesh sampling the Brillouin zone of a fcc conventional unit cell. All simulation results exclude the contact resistances (see details in Appendix~\ref{app:Rcontact}).

NEMD simulations are performed using the LAMMPS software with a domain size of $10\times10\times302$ conventional unit cells.\cite{Plimpton1995} One atomic layer at each end of the domain is set as a wall, and periodic boundary conditions are imposed along x and y directions. Langevin thermostat is used to add heat from the left side and remove heat from the right side with 2 fs time step.  The bath at each side is 50 unit cells thick, and the bath temperature is maintained at  $\textbf{T}_{bath} = (1\pm0.1)\textbf{T}$ with a time constant of 1.07 ps. Such thermostat setup ensures sufficient phonon-phonon scattering to prevent potential size effects at low temperatures. Following previous work \cite{Polanco2017,Rouzbeh2019}, comprehensive tests of size effects have been performed. The results for system sizes and temperatures relevant to the present work are summarized in in Table~\ref{size_effect_test}, and no significant impact on the interfacial thermal conductance from those factors was observed. 

\renewcommand{\thetable}{S\arabic{table}}
\setcounter{table}{0}
\begin{table}[htp]
\caption{Size effect tests for NEMD simulations on the thermal conductance of a bridged interface (MWm$^{-2}$K$^{-1}$). All these tests are done on a $L$=20 u.c. system using 5 independent simulations.}
\centering
%\begin{adjustbox}{width=\columnwidth}
%\begin{tabular}{b{.08\textwidth}b{.1\textwidth}b{.1\textwidth}b{.1\textwidth}b{.1\textwidth}} %for 2 columns
\begin{adjustbox}{width=0.6\columnwidth}
\begin{tabular}{b{.15\textwidth}b{.15\textwidth}b{.15\textwidth}b{.15\textwidth}b{.15\textwidth}} %for 1 column
 \hline
 \hline
 \\
 \multicolumn{5}{c}{$G$ (MWm$^{-2}$K$^{-1}$)}\\\\
 \hline
 \\
Size (u.c.) & 90 & 120 & 240 & 300 \\\\
 \hline
\\
 \textbf{T} = 2 K & 84.39$\pm$0.53 & 85.18$\pm$0.36 & 86.71$\pm$0.21 & 86.66$\pm$0.63 \\\\

 \textbf{T} = 30 K  & 124.98$\pm$0.38 & 126.24$\pm$0.49 & 132.63$\pm$0.73 & 131.24$\pm$1.96 \\\\

 \hline
 \hline
\end{tabular}
\end{adjustbox}
\label{size_effect_test}
\end{table}

Thermal expansion of the system was also taken into consideration. To find the temperature dependence of the lattice constant $ a(\textbf{T})$, we fitted it to the following function using the isothermal-isobaric ensemble (NPT) under zero pressures:
\begin{equation}
    a(\textbf{T}) = 5.2222 + 0.0004\textbf{T} +10^{-6} \textbf{T}^2 - 4\times10^{-9} \textbf{T}^3 \text{\AA}.
\end{equation}
The thermal conductances reported in this paper are the average of 5 sets of simulations with randomly generated initial atomic velocities. The conductance is computed by dividing the heat flux over the temperature drop across the bridge interface ($\Delta\textbf{T}_i$ in Fig.~\ref{fig_appB_Tfile}). The temperatures at the edges of the contacts used to calculate $\Delta\textbf{T}_i$ result from a linear extrapolation of the temperature profile within each contact. 

\section{Temperature profile and contact resistance}
\label{app:Rcontact}
\begin{figure}[ht]
	\centering
	\includegraphics[width=78mm]{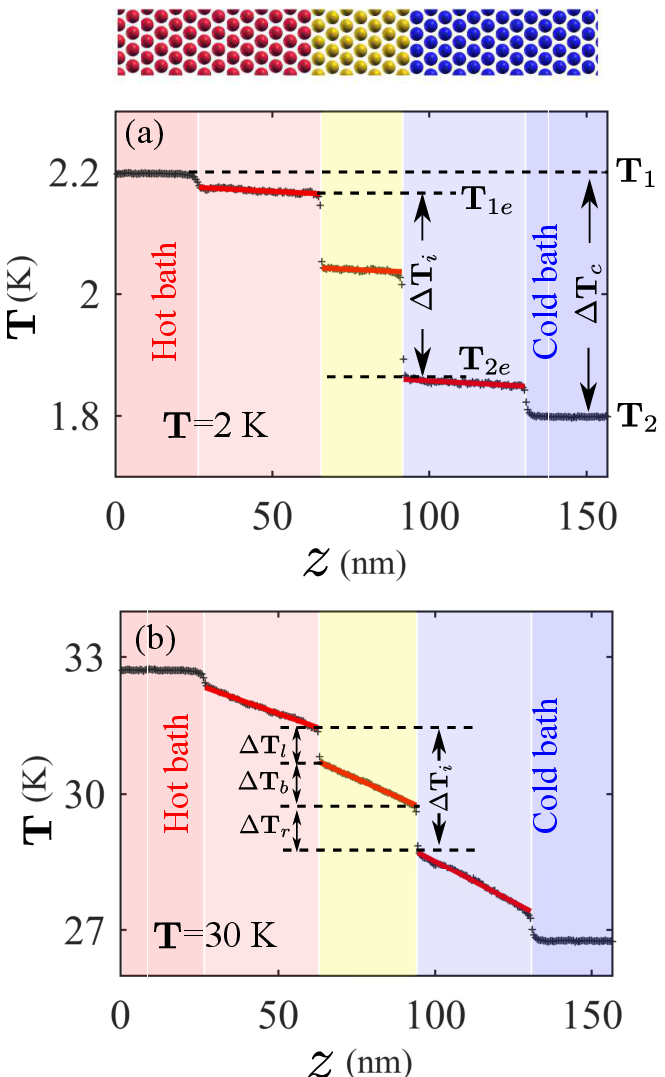}
	\caption{(a) The temperature profile at \textbf{T}=2 K when $L$=60 u.c.. $\Delta\textbf{T}_i$ and $\Delta\textbf{T}_c$ are temperature differences at the interface and between the bath contacts respectively. (b) The temperature profile at \textbf{T}=30 K when $L$=50 u.c.;$\Delta\textbf{T}_l$ and $\Delta\textbf{T}_r$ are the temperature drops at the left and right boundary respectively, $\Delta\textbf{T}_b$ is the temperature drop within the bridge layer and $\Delta\textbf{T}_i$ is the total temperature drop at the bridged interface. }
	\label{fig_appB_Tfile}
\end{figure}

 \begin{figure}[ht]
	\centering
	\includegraphics[width=78mm]{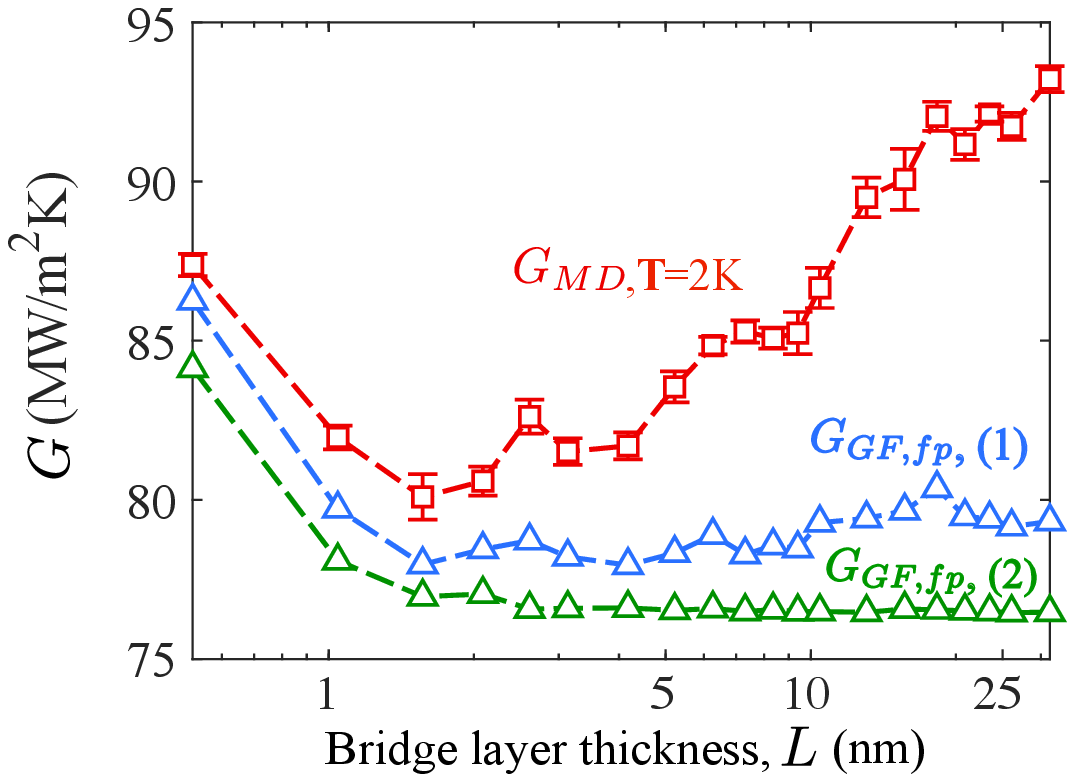}
	\caption{The interfacial thermal conductance in the harmonic limit by method (a) ($G_{GF,fp},(1)$) and method (b) ($G_{GF,fp},(2)$) in Appendix~\ref{app:Rcontact}, compared to the conductance with weak anharmonicity $G_{MD}$, \textbf{T}=2 K.}
	\label{fig_appB_G}
\end{figure}
Thermal conductance calculated using the Landauer formula (Eq.~\ref{equG}) yields a two-probe measurement of conductance, which is the heat flux over the temperature difference between the baths (shown as $\Delta\textbf{T}_c$ in Fig.~\ref{fig_appB_Tfile}(a)). To convert this value to a four-probe measurement of conductance using the temperature difference immediately at the interface ($\Delta\textbf{T}_i$ in Fig.~\ref{fig_appB_Tfile}(a)), we deduced the contact resistances following two approaches:

(a) To fairly compare calculations from NEGF with those from NEMD under weak anharmonicity ({\bf{T}}=2 K), we combined the temperature differences from NEMD simulations with Eq.~\ref{equG_T}. Thus, the four-probe conductance inferred from NEGF is given by $G_{GF,fp}=G_{GF}\frac{\Delta \textbf{T}_c}{\Delta \textbf{T}_i}$. The corresponding four-probe conductances $G_{GF,fp}$ are shown as blue triangular symbols in Fig.~\ref{fig_G}(b) and Fig.~\ref{fig_appB_G}.

(b) Instead of using the temperature differences obtained from NEMD simulations, we can use the temperature differences in Green's function simulations. This requires assigning a temperature to the non-equilibrium distributions between the baths ($\textbf{T}_{1e}$ and $\textbf{T}_{2e}$ shown in Fig.~\ref{fig_appB_Tfile}(a)). Under the equilibrium assumption \cite{Tian2012, Tian2014}, $\textbf{T}_{1e}$ and $\textbf{T}_{2e}$ can be expressed as $\textbf{T}_{1e}=\textbf{T}_1+(\textbf{T}_2-\textbf{T}_1)G_{GF}/(2G_1)$ and $\textbf{T}_{2e}=\textbf{T}_2-(\textbf{T}_2-\textbf{T}_1)G_{GF}/(2G_2)$, where $G_{GF}$ is the two-probe conductance for the whole system and, $G_1$ and $G_2$ are the conductances of the pure contact materials. As a result, the four-probe conductance can be written as \cite{Polanco2017,Tian2012, Tian2014}: 
\begin{equation}
    G_{GF,fp}=G_{GF}\times\frac{1}{1-\frac{1}{2}[\frac{G_{GF}}{G_1}+\frac{G_{GF}}{G_2}]}
\end{equation}
 The four-probe conductances $G_{GF,fp}$ calculated by this method are shown as green triangular symbols in Fig.~\ref{fig_appB_G}.

\section{Normal Mode Decomposition Technique}
\label{app:NMD}

The bulk phonon mean free path $\lambda_b$ is obtained as the product of group velocity and phonon lifetime: $\lambda_b({^{\bm{\kappa}}_{\nu}};\omega)=|v({^{\bm{\kappa}}_{\nu}};\omega)|\tau({^{\bm{\kappa}}_{\nu}};\omega)$. The phonon lifetime is calculated using the normal mode decomposition method~\cite{McGaughey2014}. Atomic velocities are projected onto eigenvectors corresponding to normal modes $({^{\bm{\kappa}}_{\nu}};\omega)$ of the bulk crystal based on lattice dynamics. Phonon lifetimes are then calculated by fitting to the decay function of the total energies. 

From ref.~\cite{McGaughey2014}, the normal mode coordinate $q(^{\bm{\kappa}}_{\nu};t)$ and its derivative with time $\Dot{q}(^{\bm{\kappa}}_{\nu};t)$ for the $b^{th}$ atom in the $l^{th}$ unit cell can be expressed as:
\begin{equation}
    q(^{\bm{\kappa}}_{\nu};t) = \sum_{b,l} \Big(\frac{m_b}{N}\Big)^{1/2} \exp{[i\bm{\kappa} \cdot \bm{r}_0(^l_0)]} \bm{e}^{*}_{b}(^{\bm{\kappa}}_{\nu}) \cdot \bm{u}(^l_b;t)
\end{equation}
and

\begin{equation}
    \Dot{q}(^{\bm{\kappa}}_{\nu};t) = \sum_{b,l} \Big(\frac{m_b}{N}\Big)^{1/2} \exp{[i\bm{\kappa} \cdot \bm{r}_0(^l_0)]} \bm{e}^{*}_{b}(^{\bm{\kappa}}_{\nu}) \cdot \bm{\Dot{u}}(^l_b;t)
\end{equation}
where $m_b$ is the mass of the atom b, $\bm{\kappa}$ and $\nu$ correspond to the wavevector and the polarization, and $\bm{r}_0(^l_0)$ is the equilibrium position of the $l^{th}$ unit cell.

The potential and kinetic energies of the normal mode are
\begin{equation}
    U(^{\bm{\kappa}}_{\nu};t) = \frac{1}{2} {\omega(_{\nu}^{\bm{\kappa}})}^2 q^* (^{\bm{\kappa}}_{\nu};t) q(^{\bm{\kappa}}_{\nu};t)
\end{equation}
and
\begin{equation}
    T(^{\bm{\kappa}}_{\nu};t) = \frac{1}{2} {\Dot{q}}^* (^{\bm{\kappa}}_{\nu};t) \Dot{q}(^{\bm{\kappa}}_{\nu};t)
\end{equation}

Total energy of the normal mode, as the sum of kinetic and potential energy of the normal mode, can be expressed as:
\begin{equation}
    \frac{\langle E(^{\bm{\kappa}}_{\nu};t) E(^{\bm{\kappa}}_{\nu};0)\rangle}{\langle E(^{\bm{\kappa}}_{\nu};0) E(^{\bm{\kappa}}_{\nu};0)\rangle} = \text{exp}[-2\Gamma(^{\bm{\kappa}}_{\nu})t]
\end{equation}
where $\Gamma(^{\bm{\kappa}}_{\nu})$ is the phonon linewidth which equals to $1/[2\tau(^{\bm{\kappa}}_{\nu})]$. The phonon lifetimes then can be calculated by fitting the normalized autocorrelation of the mode total energy to an exponential decay. 

To obtain the frequency dependent phonon lifetimes, we have:

\begin{equation}
    \langle T(^{\bm{\kappa}}_{\nu}) \rangle = \lim_{\tau_0\to\infty} \frac{1}{2\tau_0} \int_{0}^{\tau_0} {\Dot{q}}^* (^{\bm{\kappa}}_{\nu};t) \Dot{q}(^{\bm{\kappa}}_{\nu};t) dt.
    \label{eq:expectedT}
\end{equation}

Transforming Eqn.~\ref{eq:expectedT} into the frequency domain, we have

\begin{equation}
    \langle T(^{\bm{\kappa}}_{\nu};\omega) \rangle = \lim_{\tau_0\to\infty} \frac{1}{2\tau_0} \Big| \frac{1}{\sqrt{2\pi}} \int_{0}^{\tau_0} \Dot{q}(^{\bm{\kappa}}_{\nu};t) \text{exp} (-i \omega t) dt\Big|^2.
    \label{eq:freqT}
\end{equation}

The phonon frequency and linewidth can be extracted by fitting Eqn.~\ref{eq:freqT} to a Lorentzian function.

\section{Additional figures}
\label{app:figures}
\begin{figure}[ht]
	\centering
	\includegraphics[width=80mm]{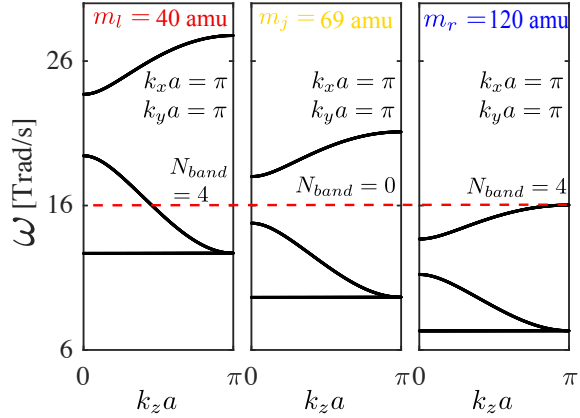}
	\caption{The projected phonon dispersion of left (red), right (blue) contact materials and in the bridge (yellow) layer when {\boldmath$\kappa^t_\perp$}$=(\pi/a,\pi/a)$. In frequency range $\omega\in(15\sim16)$ Trad s$^{-1}$, the number of phonon bands in the contacts is 4, while in the bridge layer is none. Thus non-zero $M\bar{T}$ in this range is due to phonon tunneling. There are four atoms in the fcc conventional unit cell, so each line represents 4 degenerate bands.}
	\label{projected_dispersion}
\end{figure}

\end{document}